\documentclass[runningheads]{llncs}

\usepackage{graphicx}

\begin{document}
\title{Weakly Supervised PET Tumor Detection Using Class Response}

\author{Amine Amyar\inst{1,2}\and
Romain Modzelewski\inst{2,3} \and
Pierre Vera\inst{2,3} \and
Vincent Morard\inst{1} \and 
Su Ruan\inst{2}}
\authorrunning{A. AMYAR et al.}

\institute{General Electric Healthcare, Buc, France \and
LITIS -  EA4108 - Quantif, University of Rouen, Rouen, France  \and
Nuclear Medicine Department, Henri Becquerel Center, Rouen, France}

\maketitle              
\begin{abstract}
One of the most challenges in medical imaging is the lack of data and annotated data. It is proven that classical segmentation methods such as U-NET are useful but still limited due to the lack of annotated data. Using a weakly supervised learning is a promising way to address this problem, however, it is challenging to train one model to detect and locate efficiently different type of lesions due to the huge variation in images. In this paper, we present a novel approach to locate different type of lesions in positron emission tomography (PET) images using only a class label at the image-level. First, a simple convolutional neural network classifier is trained to predict the type of cancer on two 2D MIP images. Then, a pseudo-localization of the tumor is generated using class activation maps, back-propagated and corrected in a multitask learning approach with prior knowledge, resulting in a tumor detection mask. Finally, we use the mask generated from the two 2D images to detect the tumor in the 3D image. The advantage of our proposed method consists of detecting the whole tumor volume in 3D images, using only two 2D images of PET image, and showing a very promising results. It can be used as a tool to locate very efficiently tumors in a PET scan, which is a time-consuming task for physicians. In addition, we show that our proposed method can be used to conduct a radiomics study with state of the art results.

\end{abstract}

\section{Introduction}

To better appreciate the volume of interest in oncological radiotherapy and also the biological component of a tumor, radiomics was proposed as a field of study that make use of images \cite{gillies2016radiomics}. Radiomics allows from an initial  positron emission tomography exam (PET) the prediction of the survival of a patient and the response to radio-chemotherapy treatment, and therefore to help to personalize treatment \cite{amyar20193}. The first step in a radiomics analysis is to localize tumor region for which radiomcis features can be extracted. Deep learning methods are a very promising tool for the automatic detection of lesions in PET images, but due to their data-hungry nature, they require very large amounts of annotated images, that are hard to get in medical image field. Most of segmentation methods use large annotated databases, however, annotating pixel-level tumor requires highly trainable pyhiscians and also is time-consuming. Moreover, physicians annotations can be subjective. In contrast, image-level labels indicating the presence or not of a lesion, or the type of cancer are much cheaper and can be easily obtained. This motivate the use of weakly supervised learning (WKL) approaches, where image-level information can be used to train a convloutional neural networks classifier (CNNs) to predict the class label, and with appropriate transformation to detect and localize tumors.

CNNs have yielded impressive results for a wide range of visual recognition tasks in general \cite{krizhevsky2012imagenet,girshick2014rich} and notably in medical imaging \cite{rajpurkar2017chexnet,hannun2019cardiologist}. To better understand CNNs, several works tried to visualize their internal representation \cite{mahendran2015understanding,zeiler2014visualizing,zhou2014object}. For instance, Zeiler et al \cite{zeiler2014visualizing} use deconvolutional networks to visualize activation. Different works have shown that even if a CNN is trained to classify images, it can be used to localize objects at the same time \cite{bazzani2016self,cinbis2016weakly}. Zhou et al \cite{zhou2014object} demonstrated that CNNs can recognize objects while being trained for scene recognition, and that the same network can perform both image recognition and object localization in a single training. They have shown that  convolutional units of different CNNs layers can behave as object detectors despite the lack of object localization. The deep features maps can be aggregated to extract class-aware visual evidence \cite{zhou2016learning}. However, when using  fully connected layers for classification, the ability to locate objects in convolutional layers is lost. Different studies tried to solve this problem by using a fully convolutional neural networks (FCNs) such as Network in Network (NN) \cite{lin2013network} and GoogLeNet \cite{szegedy2015going}. Typically, conventional CNNs are first converted to FCNs to produce class response maps in a single forward pass. Although image-level class labels indicate only the existence of objects classes, they can be used to meaningful indices for image segmentation, called Class Attention Maps (CAMs) \cite{selvaraju2017grad,zhou2016learning,amyar2019contribution}. These class response maps can indicate discriminant regions in the image that allow a CNN to identify an image class, however, can not distinguish different object present in the image, which make it hard to generalize to instance-level segmentation
 \cite{he2017mask} to predict precise masks at pixel-level.  Moreover, due to its resolution, CAMs cannot be directly used as supervision for pixel segmentation in PET images since they cannot distinguish between physiological fluorodeoxyglucose uptakes (normal fixation/no tumor) and pathological uptakes (tumor), see Fig1. The major concern with this method in PET images, is the difficulty to obtain a good peak corresponding to the tumor region only, because some other regions in the image can be identified due to strong gradients.

 In this work, we tackle the challenging problem of training CNNs with image-level labels for pixel-level tumor segmentation. We show that, by using a CNN architecture with appropriate modifications, we can transform the classification task to tumor localization in PET images. Specifically, we present a new method for learning pixel-level segmentation with image-level class labels, by using class response to enable a CNN for pixel mask extraction.

\begin{figure*}

		\centering

        \includegraphics[height=4cm, width=10cm]{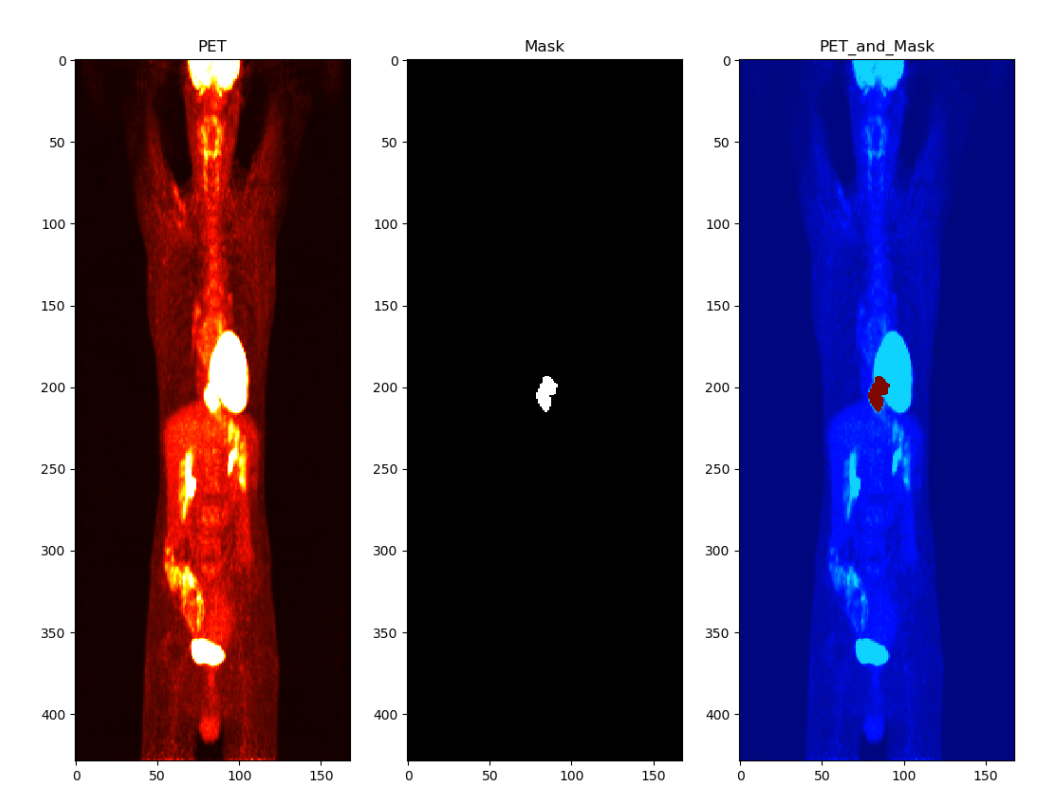}
		\caption{An example of a PET image with oesophagus cancer. It is not straightforward to learn segmentation models that can handle the difference between tumor fixation and a normal fixation in a PET image.}

  \label{fig::cancer1}
\end{figure*}

Ahn et al \cite{ahn2019weakly} presented a new approach for instance segmentation using only image-level class as label. They trained an image classifier model, and by identifying seed areas of object from attention maps, a pseudo instance segmentation labels were generated, then, propagated to discover the entire object areas with precise boundaries. Zhou et al reported that local maximums in a class response map corresponds to strong visual cues residing inside each instance \cite{zhou2018weakly}. They created a novel architecture based on peak class response for instance segmentation using only the image-level label. First, a peak from a class response map are stimulated , then, back-propagated and mapped to highly informative regions of each object instance, such as instance boundaries. 

\subsubsection{Contribution:} Using only image-level labels to segment pixel-level image remain unexplored in PET images. We propose a novel WSL-based approach to fully identify and locate tumor in 3D PET images using only two 2D images for face and profile views, the type of cancer, and simply one pixel at the center of tumor. We propose a new approach to learn pixel-level mask mapped from image-level label, which tackle the issues when using CAMs alone. Our model outperforms state-of-the-art methods trained with same supervision such as CAMs, or trained in fully supervised approach for image segmentation such as U-NET. 
\section{Method}
We use maximum intensity projection (MIP), which is a 2D image that represent 3D image for fast interpretation. The strategy to use 2d image allows to speed up the localisation, because a 3D activation map is time consuming and can be hard to train with limited resources. 
The core of our method is to implement a new class activation maps (CAMs) to locate the tumor and a new loss function to improve the precision of localisation. First, for each patient data we define one central point at the center of the tumor, that is considered as prior knowledge. Then, we define the new loss function based on the distance between the generated CAM and the central point, and the accuracy to classify the type of lesion. To that end, an 8 layers CNN was created to learn image-level labels and to generate an improved CAMs to locate correctly the lesions. After each feed forward of a mini batch of 8 images, a probability of belonging to a class is generated and a binary cross-entropy loss function is calculated (loss1). A CAM is generated for each image and a second loss for tumor localization is calculated to obtain a distance between the CAM and the central point in the tumor (loss2). Then, the back-propagation is performed in respect to both loss1 and loss2 to update the weights, see Fig2.  
\subsection{New Design of Class Activation Map (CAM)} 
CAMs are a key part in our framework.
They are used to recover the entire tumor area in a PET image by identifying pathological (tumor) fixations, and differentiate it from physiological (normal) fixations, which is done by identifying inter-pixel relations. To generate CAMs for training images, we adopt the method of \cite{zhou2018weakly} using an image classification CNN with global average pooling followed by a classification layer. Given an image, the CAM of ground truth class c is computed by
\begin{equation}
	M\textsubscript{c}(X) = \frac{W\textsubscript{c}\textsuperscript{T} f(x)}{max\textsubscript{x} W\textsubscript{c}\textsuperscript{T} f(x)'} \\
	\label{eq::CAM}
\end{equation}
where \textit{f} is the feature map from the last convolutional layer and W is the weights matrix corresponding to the class c. We backpropagate then the CAMs generated to retrieve indices of tumor fixations on the PET images, and differentiate tumor from normal fixation using a new distance loss function.
\subsection{New Loss Function} 
We introduce a novel loss function to prevent CAMs from further resolution drop. A large loss indicates that the networks current representation does not accurately capture the lesion's visual patterns, and provide an additional mechanism for self-improvement through back-propagation. As a result, the width and heigh of CAMs are 1/16 of those of the input image.
The resulting architecture is a novel convolutional neural network with attention feedback, with an improved localisation capability.

\section{Experiments}
\begin{figure*}

		\centering

        \includegraphics[height=5cm, width=12cm]{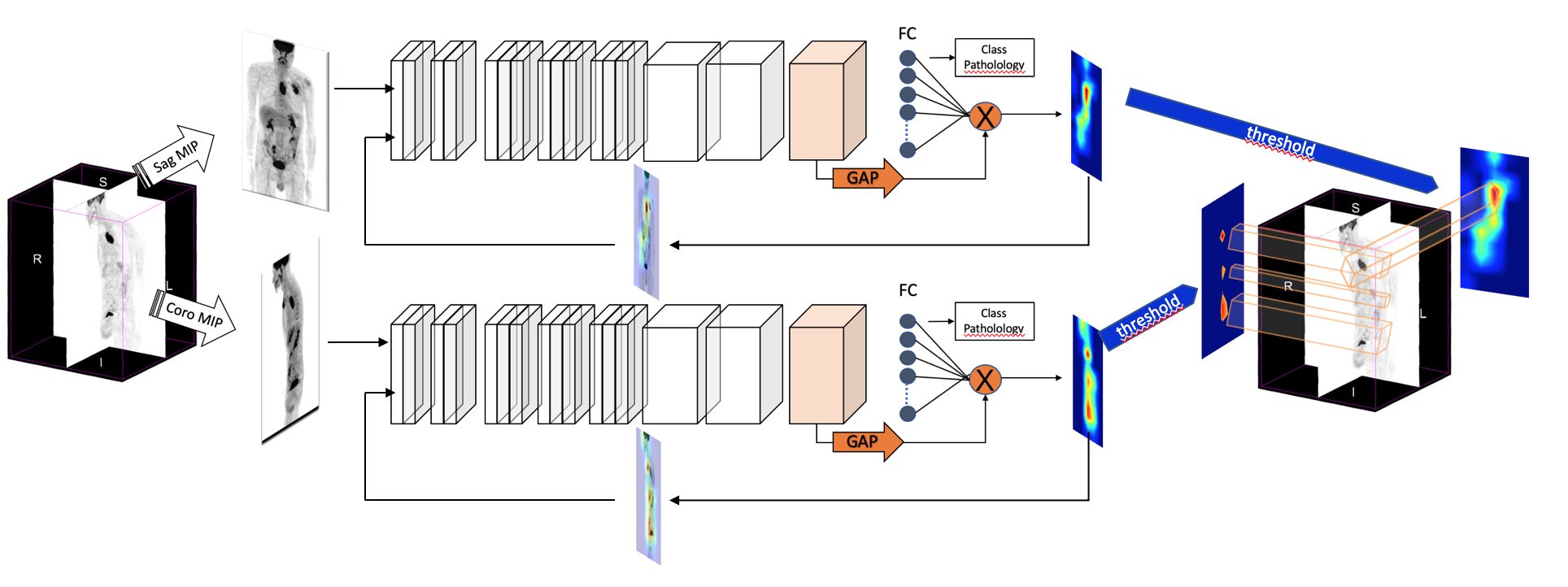}
		\caption{Our proposed architecture. The neural network learn to classify the type of cancer on two 2D MIP images (sagittal and coronal). The heatmap generated is backpropagated and corrected to identify accurately tumor regions. }

  \label{fig::cancer2}
\end{figure*}

\subsection{Dataset}
Our experiments were  run  on  195  PET  images  with  lung  (98)  and  oesophagus  (97) cancer. Patients underwent wholebody FDG PET with a CT (baseline PET), at the initial stage of the pathology and before any treatment. The PET/CT data were acquired on the same device, and with the same acquisition and reconstruction procedure used in routine care. The reconstructed exam voxel size was $4.06\times4.06\times2.0$ mm\textsuperscript{3} and were spatially normalized by re-sampling all the dataset to an isotropic resolution of $2\times2\times2$ mm\textsuperscript{3} using the k-nearest neighbor interpolation algorithm.

\subsection{Setup}
We firstly generated maximum intensity projection (MIP) for face view and for side view. MIP is a 2D image that summarize 3D images for fast interpretation. Tumor gray level intensities were normalized to SUV level between [0 30] and translated between [0 1]  to be used in CNN architecture. 
The neural network is trained to classify the type of cancer(oesophagus vs lung cancers. For each mini-batch, CAMs are generated and backpropagated and corrected via a distance function, to differentiate tumor regions from normal regions. Then, the two resulted corrected CAMs, for face and profile view are combined to retrieve the tumor in the 3D image.
\subsection{Implementation}
The model was implemented using python with pytroch deep learning library, and trained for 2 days on nvidia p6000 quadro gpu with 24gb.

\section{Results}
Table 1 shows results for tumor detection and radiomics analysis for Oesophagus cancer. Different methods were compared to our proposed model: U-net, 3D-rpet-net \cite{amyar20193}, CAMs without distance function, CAMs with FCN network. Radiomics analysis was based on the prediction of treatment response to radio chemotherapy.
Table 2 shows the same results for lung cancer, but radiomics analysis was conducted for the prediction of 3 years survival. 
\begin{table}
\caption{Segmentation and radiomics results for oesophagus cancer with state of the arts methods.}\label{tab1}
\begin{tabular}{|l|l|l|l|l|l|}
\hline
Method &  Dice & Accuracy & Sensibility & Specificity & AUC \\
\hline

U-NET & 0.42$\pm$0.16 & / & / & / & /\\
CAMs alone & 0.53$\pm$0.17 & 0.57$\pm$0.03 & 0.61$\pm$0.28 & 0.56$\pm$0.24 & 0.53$\pm$0.26\\
CAMs \& FCNs & 0.66$\pm$0.07 & 0.57$\pm$0.04 & 0.69$\pm$0.21 & 0.47$\pm$0.22 & 0.51$\pm$0.24\\
3d-rpet-Net & / & \textbf{0.72$\pm$0.08} & 0.79$\pm$0.17 &\textbf{0.62$\pm$0.21} & \textbf{0.70$\pm$0.04}\\
\textbf{Our} & \textbf{0.73$\pm$0.09} & 0.69$\pm$0.04 & \textbf{0.80$\pm$0.14} & 0.59$\pm$0.26 & 0.67$\pm$0.08\\
\hline
\end{tabular}
\end{table}

\begin{table}
\caption{Segmentation and radiomics results for lung cancer with state of the art methods methods.}\label{tab2}
\begin{tabular}{|l|l|l|l|l|l|}
\hline
Method &  Dice & Accuracy & Sensibility & Specificity & AUC \\
\hline
U-NET & 0.57$\pm$0.19 & / & / & / & /\\
CAMs alone & 0.63$\pm$0.14 & 0.61$\pm$0.07 & 0.59$\pm$0.21 & 0.57$\pm$0.15 & 0.55$\pm$0.24\\
CAMs \& FCNs & 0.73$\pm$0.12 & 0.59$\pm$0.07 & 0.63$\pm$0.12 & 0.57$\pm$0.19 & 0.57$\pm$0.17\\
3d-rpet-Net & / & \textbf{0.68$\pm$0.17} & \textbf{0.72$\pm$0.09} & 0.54$\pm$0.07 & \textbf{0.61$\pm$0.03}\\
\textbf{Our} & \textbf{0.77$\pm$0.07} & 0.65$\pm$0.05 & 0.65$\pm$0.18 & 0\textbf{0.58$\pm$0.15} & 0.59$\pm$0.04\\
\hline
\end{tabular}
\end{table}

All the methods were compared based on the ability to detect accurately the tumor and to conduct a radiomics analysis (response to treatment for oesophagus cancer and 3 years survival for lung cancer), except for 3d-rpet-net, which uses manual annotation to for radiomics, and for U-NET which is used to segmentation only. The comparison is based on accuracy, sensibility, specificity and the area under the ROC curve. The results were obtained using a 5 fold  cross-validation.
Best results for segmentation were obtained using our proposed model for both lung and oesophagus cancer. For radiomics, 3d-rpet-Net was not statistically significantly different from our model (p=0.59) for oesophagus and (p=0.63) for lung. Our model tend to have a better sensibility for oesophagus and a better specificity for lung cancer with no significant differences.
\section{Discussion \& Conclusion}

In this study, a new weakly supervised learning model was developed to localize lung and oesophagus tumors in PET images. It utilizes two fundamental components: a new class activation map to locate the tumor and a new loss function to improve localisation capability. The model could detect tumors with better accuracy compared to fully supervised models such as U-NET, or CAMs alone and CAMs plus FCNs. While 3d-rpet-net is showing slightly better results than our model in radiomics analysis, it is based in manual pixel-level annotations of tumor, which  requires a physician expert and also is time consuming. For dice coefficient: our model outperformed other methods, especially U-NET, with big margin.
The results for lung cancer were better than oesophagus cancer for tumor detection but inferior for radiomics analysis. This can be explained due to the the complexity of detecting esophageal tumor compared to lung tumor (see Fig1.), which results in better results for tumor localization. However, since the prediction of survival is harder then the prediction of the response of treatment, the performance of both 3d-rpet-Net and our model were somewhat lower.

By detecting the tumor in two 2D MIP images for face and profile views, we can obtain x,y and z coordinates to segment the 3D image. The segmentation in the 3D images were used to conduct a radiomics analysis with state-of-the-art results. This simple and yet powerful technique, can be integrated in future workflow/software dedicated to automatic analysis of PET exams to conduct radiomics analysis.

\bibliographystyle{splncs04}
\bibliography{Main}

\begin{thebibliography}{10}
\providecommand{\url}[1]{\texttt{#1}}
\providecommand{\urlprefix}{URL }
\providecommand{\doi}[1]{https://doi.org/#1}

\bibitem{ahn2019weakly}
Ahn, J., Cho, S., Kwak, S.: Weakly supervised learning of instance segmentation
  with inter-pixel relations. In: Proceedings of the IEEE Conference on
  Computer Vision and Pattern Recognition. pp. 2209--2218 (2019)

\bibitem{amyar20193}
Amyar, A., Ruan, S., Gardin, I., Chatelain, C., Decazes, P., Modzelewski, R.:
  3-d rpet-net: development of a 3-d pet imaging convolutional neural network
  for radiomics analysis and outcome prediction. IEEE Transactions on Radiation
  and Plasma Medical Sciences  \textbf{3}(2),  225--231 (2019)

\bibitem{amyar2019contribution}
Amyar, A., Decazes, P., Ruan, S., Modzelewski, R.: Contribution of class
  activation map on wb pet deep features for primary tumour classification.
  Journal of Nuclear Medicine  \textbf{60}(supplement 1),  1212--1212 (2019)

\bibitem{bazzani2016self}
Bazzani, L., Bergamo, A., Anguelov, D., Torresani, L.: Self-taught object
  localization with deep networks. In: 2016 IEEE winter conference on
  applications of computer vision (WACV). pp.~1--9. IEEE (2016)

\bibitem{cinbis2016weakly}
Cinbis, R.G., Verbeek, J., Schmid, C.: Weakly supervised object localization
  with multi-fold multiple instance learning. IEEE transactions on pattern
  analysis and machine intelligence  \textbf{39}(1),  189--203 (2016)

\bibitem{gillies2016radiomics}
Gillies, R.J., Kinahan, P.E., Hricak, H.: Radiomics: images are more than
  pictures, they are data. Radiology  \textbf{278}(2),  563--577 (2016)

\bibitem{girshick2014rich}
Girshick, R., Donahue, J., Darrell, T., Malik, J.: Rich feature hierarchies for
  accurate object detection and semantic segmentation. In: Proceedings of the
  IEEE conference on computer vision and pattern recognition. pp. 580--587
  (2014)

\bibitem{hannun2019cardiologist}
Hannun, A.Y., Rajpurkar, P., Haghpanahi, M., Tison, G.H., Bourn, C., Turakhia,
  M.P., Ng, A.Y.: Cardiologist-level arrhythmia detection and classification in
  ambulatory electrocardiograms using a deep neural network. Nature medicine
  \textbf{25}(1), ~65 (2019)

\bibitem{he2017mask}
He, K., Gkioxari, G., Doll{\'a}r, P., Girshick, R.: Mask r-cnn. In: Proceedings
  of the IEEE international conference on computer vision. pp. 2961--2969
  (2017)

\bibitem{krizhevsky2012imagenet}
Krizhevsky, A., Sutskever, I., Hinton, G.E.: Imagenet classification with deep
  convolutional neural networks. In: Advances in neural information processing
  systems. pp. 1097--1105 (2012)

\bibitem{lin2013network}
Lin, M., Chen, Q., Yan, S.: Network in network. arXiv preprint arXiv:1312.4400
  (2013)

\bibitem{mahendran2015understanding}
Mahendran, A., Vedaldi, A.: Understanding deep image representations by
  inverting them. In: Proceedings of the IEEE conference on computer vision and
  pattern recognition. pp. 5188--5196 (2015)

\bibitem{rajpurkar2017chexnet}
Rajpurkar, P., Irvin, J., Zhu, K., Yang, B., Mehta, H., Duan, T., Ding, D.,
  Bagul, A., Langlotz, C., Shpanskaya, K., et~al.: Chexnet: Radiologist-level
  pneumonia detection on chest x-rays with deep learning. arXiv preprint
  arXiv:1711.05225  (2017)

\bibitem{selvaraju2017grad}
Selvaraju, R.R., Cogswell, M., Das, A., Vedantam, R., Parikh, D., Batra, D.:
  Grad-cam: Visual explanations from deep networks via gradient-based
  localization. In: Proceedings of the IEEE international conference on
  computer vision. pp. 618--626 (2017)

\bibitem{szegedy2015going}
Szegedy, C., Liu, W., Jia, Y., Sermanet, P., Reed, S., Anguelov, D., Erhan, D.,
  Vanhoucke, V., Rabinovich, A.: Going deeper with convolutions. In:
  Proceedings of the IEEE conference on computer vision and pattern
  recognition. pp.~1--9 (2015)

\bibitem{zeiler2014visualizing}
Zeiler, M.D., Fergus, R.: Visualizing and understanding convolutional networks.
  In: European conference on computer vision. pp. 818--833. Springer (2014)

\bibitem{zhou2014object}
Zhou, B., Khosla, A., Lapedriza, A., Oliva, A., Torralba, A.: Object detectors
  emerge in deep scene cnns. arXiv preprint arXiv:1412.6856  (2014)

\bibitem{zhou2016learning}
Zhou, B., Khosla, A., Lapedriza, A., Oliva, A., Torralba, A.: Learning deep
  features for discriminative localization. In: Proceedings of the IEEE
  conference on computer vision and pattern recognition. pp. 2921--2929 (2016)

\bibitem{zhou2018weakly}
Zhou, Y., Zhu, Y., Ye, Q., Qiu, Q., Jiao, J.: Weakly supervised instance
  segmentation using class peak response. In: Proceedings of the IEEE
  Conference on Computer Vision and Pattern Recognition. pp. 3791--3800 (2018)

\end{thebibliography}
\end{document}